\documentclass[12pt]{article}
\usepackage{epsf}
\def\be{\begin{equation}}
\def\ee{\end{equation}}
\def\bea{\begin{eqnarray}}
\def\eea{\end{eqnarray}}

\begin{document}
\begin{titlepage}
\begin{center}
{\Large \bf William I. Fine Theoretical Physics Institute \\
University of Minnesota \\}
\end{center}
\vspace{0.2in}
\begin{flushright}
FTPI-MINN-18/11 \\
UMN-TH-3721/18 \\
June 2018 \\
\end{flushright}
\vspace{0.3in}
\begin{center}
{\Large \bf Radiative and $\rho$ transitions between heavy quarkonium and isovector four-quark states.
\\}
\vspace{0.2in}
{\bf  M.B. Voloshin  \\ }
William I. Fine Theoretical Physics Institute, University of
Minnesota,\\ Minneapolis, MN 55455, USA \\
School of Physics and Astronomy, University of Minnesota, Minneapolis, MN 55455, USA \\ and \\
Institute of Theoretical and Experimental Physics, Moscow, 117218, Russia
\\[0.2in]

\end{center}

\vspace{0.2in}

\begin{abstract}
The recent observation and measurement of the decay $Z_c(3900) \to \rho \eta_c$ provides the data on the relative strength of the pion and $\rho$ coupling in the corresponding transitions between exotic resonances and pure heavy quarkonium. It is argued that using these data, the heavy quark limit for the $c$ and $b$ quarks, the heavy quark spin symmetry (HQSS) and the vector dominance for photon emission by the light quarks, one can (approximately) quantitatively estimate the rates of the radiative transitions from $\Upsilon(5S)$ to the expected $G$-odd states of molecular bottomonium $W_{bJ}$. The estimate of the cross section of the processes $e^+e^- \to \gamma W_{bJ}$ at the maximum of the $\Upsilon(5S)$ resonance comes out in the ballpark of 0.1\,pb, which sets a benchmark for a possible search for these processes at BelleII.   
\end{abstract}
\end{titlepage}

The exotic bottomonium-like resonances $Z_b(10610)$ and $Z_b(10650)$~\cite{bellez}  and the charmonium-like $Z_c(3900)$~\cite{besz39} and $Z_c(4020)$~\cite{besz40}, contain a light quark-antiquark pair in addition to a heavy $b \bar b$ or $c \bar c$ pair. In hadronic decays of these resonances to heavy quarkonium the light quark pair is emitted as a light meson. Previously only the decays of this type with emission of a pion were observed until the very recent report~\cite{besrho} of the observation of a $\rho$ meson emission in the decay $Z_c(3900) \to \rho \eta_c$, which provides the first insight in the relative strength of the conversion of the light quark pair from an exotic four-quark state to $\rho$ versus the conversion to $\pi$. The exotic states of this type, if interpreted as molecular~\cite{vo} ones, have certain (approximate) symmetries. The purpose of this paper is to use these symmetries and the measured relative strength of the $\rho$ and $\pi$ coupling to derive quantitative estimates for rates of other processes involving the heavy molecular states, including the radiative transitions from $\Upsilon(5S)$ to yet unknown isovector bottomonium-like molecules, using the model of $\rho$ dominance in the isovector electromagnetic current.  It is thus assumed throughout the present text that the $Z_b$ resonances are threshold molecular states of the $B \bar B^*$ and $B^* \bar B^*$ heavy meson pairs~\cite{bgmmv} and that the $Z_c(3900)$ resonance is a threshold state in the $D \bar D^*$ channel~\cite{mv13}, i.e. it is the charmonium-like sector analog of the bottomonium-like $Z_b(10610)$ resonance.

The structure of the bottomonium-like states $Z_b$ with respect to the total spin of the heavy quark-antiquark pair and the spin of the light degrees of freedom appears to be very close to that for non-interacting meson-antimeson pairs in the $S$ wave state with the corresponding quantum numbers. Unlike  the (approximate) conservation of the spin of the heavy quarks, expected from HQSS, the preservation by the interaction of the spin of the light quarks is surprising and is indicated~\cite{mv16,wbfhnw} by the experimentally observed~\cite{belle15} strong suppression of the decay $Z_b(10650) \to B \bar B^* + {\rm c.c.}$. The spin symmetry then implies existence of additional near-threshold resonances. Indeed,  by combining the spins of heavy and light quarks in the same isovector $S$-wave state as the $Z_b$ resonances, whose quantum numbers are $I^G(J^P)=1^+(1^+)$ one finds four additional $I^G=1^-$ states~\cite{mv11}: one scalar $W_{b0}$ at the threshold of $B \bar B$, one axial $W_{b1}$  made from $B \bar B^{*} + B^{*} \bar B$, and another scalar $W_{b0}^{\,'}$ and a tensor $W_{b2}$ at the threshold of $B^{*} \bar B^*$. The approximate spin independence of the interaction strongly suggests that all these four states should exist as resonances near the corresponding thresholds, similar to the $Z_b(10610)$ and $Z_b(10650)$ peaks. Unlike the $Z_b$ states the expected $W_{bJ}$ resonances are not allowed by the $G$ parity to emerge as a result of a one-pion transition from a $J^{PC}=1^{--}$ state produced in $e^+e^-$ annihilation. Only  transitions with emission of minimum two pions  in the $I^G=1^+$ state are possible. This channel for two pions is well known to be dominated by the $\rho$ meson, so that such transitions can have a significant rate if there is sufficient energy for emission of the $\rho$ meson. At present however there is no facility supplying such energy with a large luminosity neither in the bottomonium, nor in charmonium region~\footnote{The expectation for existence of additional near threshold resonances applies to both bottomonium-like and charmonium-like sectors. However isospin properties near charmed meson-antimeson thresholds are complicated by a significant isotopic mass differences of the $D^{(*)}$ mesons, as is well known for the resonance $X(3872)$.}. 

At energy below the threshold for emission of $\rho$ it is however possible to produce the expected $G$-odd resonances by radiative transitions~\cite{mv11}, i.e. by having the emission of a  $\rho^0$ substituted by emission of a photon from a state produced directly in $e^+e^-$ annihilation. Naturally, the rate for such transitions is suppressed by the electromagnetic constant $\alpha$, and their observation would require a significant statistics of data, which hopefully will be achievable in the imminent BelleII experiment. It can be noted that in these processes the current emitting the photon is necessarily an isovector one, $(\bar u \gamma_\mu u - \bar d \gamma_\mu d)/2$, and one can apply the $\rho$ meson vector dominance for description of the process. In other words, it is still an interaction with the $\rho$ that is underlying the radiative processes $e^+ e^- \to \Upsilon(5S) \to \gamma \, W_{bJ}$.

Proceeding with the discusssion of the transitions with emission of either a pion or $\rho$, we recall that the amplitude of the decay $Z_c \to \pi J/\psi$ has the form
\be
A_\pi = C_\pi \, (\vec Z \cdot \vec \psi) \, E_\pi
\label{api}
\ee
where $C_\pi$ is a constant, $\vec Z$ and $\vec \psi$ are the polarization amplitudes of the $Z_c$ and $J/\psi$, and the presence of the factor of the pion energy $E_\pi$ is mandated by the soft pion theorems. For the process with the $\rho$ meson, $Z_c \to \rho \eta_c$ the amplitude can be written as
\be
A_\rho= C_\rho \, \left [ E_\rho \, (\vec Z \cdot \vec \rho) - \rho_0 \, (\vec Z \cdot \vec q) \right ]~,
\label{arho}
\ee
where $\rho_\mu = (\rho_0, \vec \rho)$ is the 4-vector polarization amplitude of the $\rho$, and $E_\rho$ and $\vec q$ are its energy and momentum. The form of the amplitude (\ref{arho}) is determined by the requirement that the $\rho$ meson is emitted by a conserved isovector vector current, so that the ampplitude involves the component $\rho_{0i}$ of the tensor $\rho_{\mu \nu} = q_\mu \rho_\nu - q_\nu \rho_\mu$. One can notice that the constants $C_\rho$ and $C_\pi$ have the same dimension, and that the ratio of the decay rates is given by
\be
{\Gamma(Z_c \to \rho \eta_c) \over \Gamma(Z_c \to \pi J/\psi)} = {|C_\rho|^2 \over |C_\pi|^2 } \, { 2 \, E_\rho^2 + m_\rho^2 \over 3 \, E_\pi^2} \, {|\vec q_\rho| \over |\vec p_\pi|}~.
\label{rr}
\ee
In this expression any effects of the recoil for charmonium are neglected, since these would be of a higher order in the inverse of the heavy quark mass. 

The experimental value~\cite{besrho} of the ratio in Eq.(\ref{rr}) for $Z_c(3900)$ is $2.1\pm0.8$, which results in the estimate
\be
R \equiv {|C_\rho|^2 \over |C_\pi|^2 } = 2.9 \pm 1.1~.
\label{rc}
\ee

In the quark picture the amplitudes of hadronic transitions between the exotic four-quark resonances and heavy quarkonium with emission of a light meson are determined by two major factors: the heavy quark pair transition to a specific quarkonium state and the conversion of the light quark-antiquark pair into a meson. In the decays $Z_c \to \pi J/\psi$ and $Z_c \to \rho \eta_c$ in the heavy quark limit applied to the charmed quarks, the charmonium states are related by HQSS and their spatial wave function is the same. Thus the ratio of the heavy quarkonium overlap factors is determined by the spin structure of the heavy quark pair in the molecular state. For a pair of free mesons in the $I^G(J^P)=1^+(1^+)$ state the spin structure in terms of the total spin-parity of the heavy (H) and the light (L) degrees of freedom reads as~\cite{bgmmv}
\be 
D^* \bar D - D \bar D^*  \sim 0^-_H \otimes 1^-_L + 1^-_H \otimes 0^-_L ~,
\label{zs}
\ee
and contains equal mixture of the ortho- ($S_H=1$) and para- ($S_H=0$) spin states of the heavy quark pair. If the threshold resonance $Z_c(3900)$ retains this spin structure, this would explain essentially equal strength of its transitions to ortho-charmonium (e.g. $J/\psi$) and to para-charmonium 
(e.g. $\eta_c$). This behavior of the discussed molecular states was first observed for the $Z_b$ resonances. Moreover, it is very likely that the actual spin structure of the $Z_b(10610)$ and $Z_b(10650)$ resonances is quite close to that of free meson pairs $B^* \bar B - B \bar B^*$ and $B^* \bar B^*$, as indicated by the observed~\cite{belle15} virtual absence of the decays of the heavier $Z_b(10650)$ resonance to the lighter pairs $B^* \bar B$ and $B \bar B^*$. The only known reason for suppression of this decay is the spin orthogonality, with the pairs $B^* \bar B - B \bar B^*$ having the spin structure described by Eq.(\ref{zs}) (naturally with the $c$ quarks replaced by the $b$) while the $Z_b(10650)$ retains the spin structure of the $B^* \bar B^*$ pair:
\be
B^* \bar B^* \sim (0^-_H \otimes 1^-_l - 1^-_H \otimes 0^-_L)/\sqrt{2}~.
\label{zps}
\ee 
It should be mentioned that at some level the spin structure of the $Z_b$ resonances is expected~\cite{mv17} to differ from that of free meson pairs, but the effects of such difference are apparently small and can be neglected in the present discussion. Moreover it is assumed here that, similarly to the $Z_b$ resonances, the spin structure of the $Z_c$ resonances can be approximated by that of the free meson pairs. Applying then Eq.(\ref{zs}) for description of the $Z_c(3900)$ resonance one concludes that the ratio of the heavy $c \bar c$ overlap factors in the amplitudes of the decays $Z_c \to \pi J/\psi$ and $Z_c \to \rho \eta_c$ is approximated by one, and that the ratio of the amplitudes is given by the conversion factors of the light quark pairs $A(1^-_L \to \rho)$ and $A(0^-_L \to \pi)$ that can be written as 
\be
A(1^-_L \to \rho) = a_\rho \, \left [ E_\rho \, (\vec \phi \cdot \vec \rho) - \rho_0 \, (\vec \phi \cdot \vec q) \right ]~,~~~~A(0^-_L \to \pi)= a_\pi \, E_\pi~,
\label{aropi}
\ee
where $\vec \phi$ is the polarization amplitude for the $1^-_L$ state and $a_\rho$, $a_\pi$ are constants whose ratio determines that of the amplitudes $C_\rho$ and $C_\pi$, $a_\rho/a_\pi = C_\rho/C_\pi$, given that the $0^-_L$ and $1^-_L$ come with the same weight in the structure (\ref{zs}).

In the picture considered here  we assume, for an estimate, that the ratio $a_\rho/a_\pi$ is universal for the discussed isovector threshold resonances. Furthermore, if the limit of heavy  mass is applied to both $c$ and $b$ quarks, this ratio can be approximated as being the same for the charmonium-like and bottomonium-like exotic states, although the absolute value of each coefficient can be different between the two systems. Then, one can estimate e.g. the rates of not yet observed decays of the $Z_b$ resonances as
\be
  {\Gamma[Z_b \to \rho \eta_b(1S)] \over \Gamma[Z_b \to \pi \Upsilon(1S)]} \approx 0.7 \, R \approx 2~,
\label{rgzb}
\ee
where any differences in the kinematics between the decays of $Z_b(10610)$ and $Z_b(10650)$ as well as between the transitions to $\Upsilon(1S)$ and $\eta_b(1S)$ are neglected, since these differences are likely smaller than the uncertainty of the estimate. Moreover these differences are formally of a higher order in the inverse of the $b$ quark mass, and it would be not justified to retain such terms within the discussed approximation. The ratio of the kinematical factors for the $\rho$ and the pion from Eq.(\ref{rr}) is however retained and results in the numerical factor 0.7 in Eq.(\ref{rgzb}).	One can notice that this factor goes to 2/3 (the fraction of `active' polarizations for the $\rho$) in the limit where both the pion and the $\rho$ masses are neglected.

The extrapolation from the observed ratio of the $\rho$ and $\pi$ emission rates in the charmonium sector to the bottomonium one is likely the biggest uncertainty in the presented here estimates. It is assumed here that the isovector threshold molecular state $Z_c(3900)$ is similar to the $Z_b$ resonances in that it closely retains the spin structure of a state of two free mesons [Eq.(\ref{zs})]. It is this assumption, based on applying the heavy mass limit to both $c$ and $b$ quarks, that is necessary to arrive at Eq.(\ref{rgzb}) from Eq.(\ref{rr}) and whose accuracy is not known at present. Clearly, it would immensely contribute to resolving this uncertainty, if the ratio of rates in Eq.(\ref{rgzb}) was measured experimentally. 

As is previously discussed, the most interesting at the moment application of the approximations considered here is a quantitative estimate of the rates of the radiative transitions from $\Upsilon(5S)$ to the hypothetical $C$-even neutral isovector $W_{bJ}$ resonances. The assumption of vector dominance for the photon emission in the $\Delta I=1$ transitions, amounts to replacing the polarization amplitude for $\rho$ with that for a photon, $\vec \epsilon$, as $\vec \rho \to e \, \vec \epsilon / g_\rho$, where $e$ is the electromagnetic constant $e^2 = 4 \pi \alpha$ and $g_\rho$ being the known constant, $g_\rho^2/4\pi \approx 2.7$. The rates of the radiative processes can be normalized to those of the observed~\cite{belle15} processes $e^+e^- \to  \Upsilon(5S) \to \pi Z_b$, whose cross section at the maximum of the $\Upsilon(5S)$ peak is in the ballpark of 10\,pb for both $Z_b$ resonances, and taking into account only one charge combination in the final state, e.g. $\pi^0 Z_b^0$~\footnote{The estimate of the $\Upsilon(5S)$ peak  cross section from the data reported in Ref.\cite{belle15} is discussed in Ref.~\cite{mv16sr}.}. In order to account for the heavy quark spin factor, we recall the spin structure of the four expected $W_{bJ}$ states~\cite{mv11}:
\bea
&& W_{b0}  \sim  B \bar B \sim {1 \over 2} \, \left ( 0^-_H \otimes 0^-_L \right ) - {\sqrt{3} \over 2} \,\left. \left ( 1^-_H \otimes 1^-_L \right ) \right |_{J=0} ~, \nonumber \\
&& W_{b1} \sim B \bar B^* + B^* \bar B \sim \left. \left ( 1^-_H \otimes 1^-_{L} \right ) \right |_{J=1}~, \nonumber \\
&& W_{b0}^{\,'} \sim \left . B^* \bar B^* \right |_{J=0} \sim {\sqrt{3} \over 2} \, \left ( 0^-_H \otimes 0^-_{L} \right ) + {1 \over 2} \,\left. \left ( 1^-_H \otimes 1^-_{L} \right ) \right |_{J=0}~, \nonumber \\
&& W_{b2} \sim \left . B^* \bar B^* \right |_{J=2} \sim \left. \left ( 1^-_H \otimes 1^-_{L} \right ) \right |_{J=2}~.
\label{ws}
\eea
Considering the $\Upsilon(5S)$ state as pure $1^-$ ortho-bottomonium, one can then determine the heavy spin projection coefficients and finally estimate the relative rates of the transitions in terms of the ratio $R$ as follows
\bea
&&{\Gamma[\Upsilon(5S) \to \gamma \, W_{b0}] \over \Gamma[\Upsilon(5S) \to \pi^0 Z_b^0(10610)]} = {1 \over 3 } \, \alpha \, {4 \pi \over g_\rho^2} \, R \,  {E_\gamma^3 \over E_\pi^2 \, p_\pi} \approx 0.5 \times 10^{-2}~, \nonumber \\
&&{\Gamma[\Upsilon(5S) \to \gamma \, W_{b1}] \over \Gamma[\Upsilon(5S) \to \pi^0 Z_b^0(10610)]} = {4 \over 3 } \, \alpha \, {4 \pi \over g_\rho^2} \, R \,  {E_\gamma \over  p_\pi} \approx 1.2 \times 10^{-2}~, \nonumber \\
&&{\Gamma[\Upsilon(5S) \to \gamma \, W_{b0}^{'}] \over \Gamma[\Upsilon(5S) \to \pi^0 Z_b^0(10650)]} = {1 \over 9 } \, \alpha \, {4 \pi \over g_\rho^2} \, R \,  {E_\gamma \over  p_\pi} \approx 0.1 \times 10^{-2}~, \nonumber \\
&&{\Gamma[\Upsilon(5S) \to \gamma \, W_{b2}] \over \Gamma[\Upsilon(5S) \to \pi^0 Z_b^0(10650)]} = {20 \over 9 } \, \alpha \, {4 \pi \over g_\rho^2} \, R \,  {E_\gamma \over  p_\pi} \approx 2.2 \times 10^{-2}~.
\label{gest}
\eea

It can be also noted that a hadronic process $\Upsilon(5S) \to \pi \pi W_{b0}$ is kinematically allowed for the lowest $W_{b0}$ resonance expected at the threshold of the pseudoscalar meson pairs $B \bar B$. However the phase space for this channel is very small. Assuming for an estimate that the pions are described by the low mass `tail' of the $\rho$ resonance one can estimate the rate for this process relative to the radiative one as
\be
{\Gamma[\Upsilon(5S) \to \pi^+ \pi^- W_{b0}] \over \Gamma[\Upsilon(5S) \to \gamma \, W_{b0}]} \approx
\left ( {g_\rho^2 \over 4 \pi} \right )^2 \, {1 \over 4 \pi \alpha m_\rho^4} \, \int q^2 \, \sqrt{1-{q^2\over \Delta^2}} \, \left ( 1- {4 m_\pi^2 \over q^2} \right )^{3/2} \, d q^2 \approx 0.012~,
\label{ppg}
\ee
with the integral over $q^2$ running between the kinematical limits, i.e. from $4 m_\pi^2$ to $\Delta^2$ with $\Delta = M[\Upsilon(5S)]- M[W_{b0}]$. 

The estimates (\ref{gest}) put the $e^+e^-$ cross section of the discussed radiative processes in the ballpark of 0.1\,pb. In lieu of a higher electron-positron collider energy these processes appear to be the only potentially accessible gateway to the $G$-odd $W_{bJ}$ molecular states in the bottomonium sector. Once produced, these states are expected~\cite{mv11} to decay into $\rho \Upsilon(1S)$ with a branching fraction likely comparable to that of the decays $Z_b \to \pi \Upsilon(1S)$, i.e. of the order of one percent. The presence of the $\Upsilon(1S)$ in the final state may significantly reduce the background in a search for the $W_{bJ}$ states in this channel. It is also clear from Eqs.(\ref{gest}) that the largest rate of the radiative transitions is expected for the tensor $W_{b2}$ molecular state at the threshold of pairs of vector mesons $B^* \bar B^*$. According to the relations (\ref{ws}) this state, as well as the axial $W_{b1}$ at the threshold of $B^* \bar B$, contains only the $1_H^-$ state of the $b \bar b$ pair and thus should decay mostly to ortho-bottomonium, which may enhance the branching fraction for the final channel $\rho \Upsilon(1S)$. In fact, for the heaviest $W_{b2}$ state there is also an option of searching for it in the decay channel $W_{b2} \to \pi^+ \pi^- \Upsilon(2S)$ with the two pions being in the low-mass tail of the $\rho$ resonance. In this case the kinematical suppression due to lack of phase space for  $\rho$ emission corresponds to a factor of about 0.1, which may be at least partially compensated by the enhancement of the transition to $\Upsilon(2S)$ in comparison with $\Upsilon(1S)$, if the decays to various states of bottomonium for the $W_{b2}$ follow a similar pattern as for the $Z_b$ resonances~\cite{belle15}~\footnote{This possibility was suggested by A.~Bondar.}.

It is clear that the estimates for the rates in Eqs.(\ref{gest}) rely on a number of approximations, including the HQSS and also a symmetry between the interactions of the light degrees of freedom in the $0_L^-$ and $1_L^-$ states. The latter symmetry is in fact a surprising logical conclusion~\cite{mv16,wbfhnw} from the data~\cite{belle15} on the decays of the $Z_b$ resonances to heavy meson pairs. It is not known whether a similar behavior takes place in the charmonium sector. At present the only justification of assuming that it does is based on applying the heavy mass limit to both the $c$ and the $b$ quarks, so that it is difficult to evaluate the uncertainty of the presented estimates. This assumption could be tested experimentally if the ratio (\ref{rgzb}) of the transition rates from one or both $Z_b$ resonances was measured. In case the actual ratio turns out to be significantly different from the numerical value in Eq.(\ref{rgzb}), the estimates (\ref{gest}) of the rates of the radiative transitions will have to be rescaled accordingly.

Finally, it should be noted that an application of the presented approach to the $C$-even threshold resonances in the charmonium sector involve considerable additional uncertainties. Indeed, there is a significant violation of the isotopic symmetry near the thresholds for the charmed meson pairs due to  large isotopic mass differences for both the pseudoscalar and the vector charmed mesons. Thus the neutral components of the isotopic triplets of the molecular states generally strongly mix with the isosinglets and can thus mix with the states of pure $c \bar c$ charmonium, which behavior is well known for the state $X(3872)$ (a pure isotriplet state would be a charmonium-like analog of the axial $W_{b1}$). For such mixed states the emission of the radiative photon in the processes $e^+ e^- \to \gamma X_c$ is determined not only by the isotriplet electromagnetic current of the light quarks, but also by the isosinglet one, including the current of the charmed quarks. Clearly, the amplitudes under these circumstances involve new effects whose estimate at present would be highly model dependent.

This work is supported in part by U.S. Department of Energy Grant No.\ DE-SC0011842.

\end{document}